\PassOptionsToPackage{square, numbers, sort}{natbib}
\documentclass[reprint,aps,prr]{revtex4-2}

\usepackage[utf8]{inputenc}
\usepackage[T1]{fontenc}
\usepackage[english]{babel}

\usepackage[fleqn]{amsmath}
\usepackage{mathtools}
\usepackage{amssymb}
\usepackage{xfrac}
\usepackage{physics}

\usepackage{placeins}
\usepackage{graphicx}
\usepackage{caption}
\usepackage[percent]{overpic}

\usepackage{booktabs}

\usepackage{tabularx}
\usepackage[graphicx]{realboxes}
\usepackage{array}

\usepackage{xcolor}

\newcommand{\ii}{\mathrm{i}}
\newcommand{\ee}{\mathrm{e}}

\usepackage{ulem}
\usepackage{siunitx}

\begin{document}

\title{Laser-Dressed States on Riemannian Manifolds: A Generalization of the Kramers-Henneberger Transformation}
\author{Hannah Bendin}\thanks{These authors contributed equally to this work.}
\author{Benjamin Schwager$^*$}
\author{Jamal Berakdar} 
\affiliation{Institut für Physik, Martin-Luther-Universität Halle-Wittenberg}
\date{\today}
\begin{abstract}
Quantum particles under geometric constraints are sensitive to the geometry and topology of the underlying space. We analytically study the laser-driven nonlinear dynamics of a quantum particle whose motion is constrained to a two-dimensional Riemannian manifold embedded in a three-dimensional hyperspace. The geometry of space results in a potential-like term that supports bound states on the manifold. In the presence of a laser field, we derive expressions for a generalized Kramers-Henneberger-type unitary transformation which is shown to be generally space- and time-dependent, and deduce a Schrödinger-like equation in the Kramers-Henneberger frame. Compared to a flat (geometrically trivial) space, new time-averaged coefficients of differential operators and operator-valued perturbation terms appear which determine the geometry-dependent laser-dressed states on Riemannian manifolds.
\end{abstract}
\maketitle


\section{Introduction}
Current coherent light sources are capable of generating short electromagnetic pulses with an electric field amplitude as large as the electrostatic fields that prevail in a typical atom \citep{RevModPhys.84.1177}. When an atom is exposed to such fields one may expect the bound electrons to be released rapidly. Instead, depending on the driving field frequency, bond stabilization against ionization is observed, meaning, an increase in the intensity does not lead to an increase in the ionization rate. One explanation of this phenomenon refers to destructive interference of electron-emission amplitudes from closely spaced states. A further possibility is the formation of laser-dressed states: The charged particle is dynamically stabilized by an effective potential resulting from the combined action of the residual potential and the laser fields \citep{Popov1999, Popov2003}, a situation which is referred to as Kramers-Henneberger (KH) stabilization \citep{Kramers1956, Henneberger1968}. Theoretically, the KH scenario is understood by switching to the accelerated frame of the charged particle which is then displaced by the dynamical amount $\boldsymbol{\alpha}(t) \propto \int_{-\infty}^t \mathbf{A}(t^\prime)\,dt^\prime$, where $\mathbf{A}$ is the laser field vector potential and $\alpha$ quantifies the quiver amplitude. This approach is useful in interpreting different aspects of strong field processes in atomic physics, such as the understanding of structural features like dichotomy \citep{Pont1988} and polychotomy \citep{Reed1991}, stabilization against ionization \citep{Volkova1997, Liang2022} or above threshold ionization \citep{Su1990}. For more details we refer to \citep{Vivirito1995, Popov2003, Morales2011, Yamanouchi2021} and the references therein.
\par
A key point of the conventional KH-approach is the independence of $\mathbf{A}$ on spatial coordinates, so $\boldsymbol{\alpha}(t)$ is readily obtained. This assumption is justified when $\mathbf{A}$ varies only weakly on the scale of the particle localization length which may not be the case when studying light-matter interaction phenomena associated with large photonic wave vectors \cite{PhysRevLett.95.043601, Bandrauk_2013}, or when $\mathbf{A}$ is spatially polarization or phase structured below the optical diffraction limit \cite{PhysRevA.102.063105}. Here, we study new aspects when the space itself is not uniform. We employ the confining potential approach (CPA) which, in essence, means that the particle motion is constrained (e.g., by a potential) to move on a submanifold of a higher-dimensional space \citep{Jensen1971, Costa1981, Brandt2017}.
\par
Within the CPA we derive an effective Schrödinger equation describing the motion of a quantum particle confined to a two-dimensional Riemannian manifold (which is embedded in a three-dimensional hyperspace) in the presence of electromagnetic fields. The effective Schrödinger equation contains terms that can be related to the geometry of the underlying space with a potential-like term acting (in our case) attractive and hence supporting bound states \citep{Atanasov2007}. 
Such geometry-induced bound states on the manifold can be spatially extended. In addition, even if $\mathbf{A}$ is homogeneous in the hyperspace, the nontrivial geometry of the manifold introduces a space-dependence to the vector potential part that triggers the dynamics on the manifold, thus effectively structuring the driving field. For a consistent description of these aspects, a formal treatment of light-matter interaction under constraints is desirable.
\par
In what follows we seek an extension of the KH formalism \citep{Henneberger1968} within the CPA by introducing a space- and time-dependent unitary transformation. We mathematically derive a modified KH-Schrödinger equation that contains a set of terms due to the nontrivial geometry of the underlying space. The analytical derivation relies on the dipole approximation and on series truncation which compromises unitarity in general, but for typical field strengths our approximations shall be reasonable. An important aspect is that the effective vector potentials uncovered here are enforced by the geometry of space and thus exhibit different properties than those appearing within the nondipolar extensions of the KH theory in Euclidean space \citep{Foerre2005, Foerre2020}.
Generally, geometry-induced effects on wave propagation under constraints have been experimentally detected \citep{Szameit2010, Onoe2012}. Possible candidates for an experimental verification of our predictions are rippled two-dimensional materials such as free-standing graphene or TMDC sheets under high-frequency, intense fields or tip-induced deformations \cite{Harats2020}.
\par
The article is organized as follows. Sec.\ \ref{sec:dimensional_reduction} recapitulates the confining potential approach and introduces the notation. Sec.\ \ref{sec:kh_unitary_transformation} discusses the extension of the KH-transformation to general Riemannian manifolds with emphasis on the geometry-induced potential-like term and additional terms resulting from the Laplace-Beltrami operator. Sec.\ \ref{sec:conclusion} summarizes the findings. Technical details are deferred to appendices. Throughout the work, we adopt the Ricci calculus with the following conventions: When the transition into local reference frames of the manifold is performed, the position coordinates are labelled by capital Greek indices $\{M; N; K; \ldots\}$. During the CPA, external variables describing tangent degrees of freedom are promoted and numerated by small Greek indices $\{ \mu ;\nu; \kappa; \ldots \}$.

\section{Dimensional Reduction} \label{sec:dimensional_reduction}
We consider a nonrelativistic spinless quantum particle  constrained to move on a noncompact two-dimensional Riemannian manifold $\left( \mathcal{M},g \right)$ which is isometrically embedded in the three-dimensional Galileian spacetime as $\mathcal{E}: \mathcal{M}\times \mathbb{R}^{1} \rightarrow \mathbb{R}^{3}\times \mathbb{R}^{1}$ (for details see e.\,g. \citep{Jost2011}). For simplicity, we assume that $\mathcal{M}$ is orientable and can be globally (meaning up to a point set of measure zero) parametrized by
\begin{align}
\begin{aligned}
    \mathcal{Y}: \mathbb{R}^{2} \supset \mathcal{Q}_{\mathrm{t}} &\rightarrow \mathcal{M}\subset \mathbb{R}^{3}\ ,\\
    \textbf{q} = \left( q^{1}, q^{2} \right)^{T} &\mapsto \mathcal{Y}(\textbf{q})\ .
\end{aligned}
\end{align}
We  define a unit normal vector field such that every position $\mathcal{X}\in\mathbb R^{3}$ in sufficiently close proximity to $\mathcal{M}$ is expressible  as $\mathcal{X}(\textbf{q}, y) = \mathcal{Y}(\textbf{q}) + y\,\textbf{n}(\textbf{q})$, where $y$ measures the displacement along the local normal direction. From this parametrization follows a canonical frame field that yields a local set of basis vectors $\{ \textbf{t}_{1}, \textbf{t}_{2}, \textbf{n} \}\subset \mathbb{R}^{3}$ suitable for the formulation of the Schrödinger equation in local coordinates. It canonically induces a Riemannian metric structure, and the Levi-Civita connection acts on the tangent bundle $T\mathcal{M}$. The electromagnetic field of the laser enters the theory as a gauge field. The covariant derivative yields the gauge covariant derivative with components $\tilde{\nabla}_{M} = \nabla_{M} - \frac{\ii Q}{\hbar} A_{M}\hat{I}$ and $\tilde{\partial}_{t} = \partial_{t} + \frac{\ii Q}{\hbar} \Phi\hat{I}$, with $Q$ is the electric charge and $\Phi$ and $\textbf{A}$ the electromagnetic scalar and vector potentials which depend on $(\textbf{q}, y, t)$.
\par
Within this framework for the light-matter coupling we consider the Schrödinger equation that is pulled back into the parameter space $\mathcal{Q}_{\mathrm{t}} \subset \mathbb{R}^{2}$. The motion is restricted to the Riemannian manifold by a confining potential $V_{\lambda}$ that has the following properties \citep{Brandt2017, Maraner1995, Schuster2003, Jalalzadeh2005, Maraner2008}: (i) $V_{\lambda}$ depends only on the normal displacement coordinate $y$; (ii) $V_{\lambda}$ has a deep minimum on $\mathcal{M}$ (for $y=0$) so that it can be expanded around $\mathcal{M}$, accounting for the uncertainty principle of Heisenberg; (iii) $V_{\lambda}$ preserves the gauge group of the resulting effective theory representing a subgroup of the isometry group of the hyperspace. The Schrödinger equation in local frames reads
\begin{align}
    \ii\hbar\,\tilde{\partial}_{t}\psi = \left[ -\frac{\hbar^{2}}{2m}\tilde{\nabla}^{M}\tilde{\nabla}_{M} + V_{\lambda}\,\hat{I} \right]\psi.
\end{align}
The kinetic term is proportional to the gauge-covariant Laplace-Beltrami operator
\begin{align}
\begin{aligned}
    \Delta = \tilde{\nabla}^{M}\tilde{\nabla}_{M} = \dfrac{1}{\sqrt{\vert G \vert}}\tilde{\nabla}_{M} \left( \sqrt{\vert G\vert} G^{MN}\tilde{\nabla}_{N} \right)\ ,
\end{aligned}
\label{eq:Laplace-Beltrami}
\end{align}
where $\vert G \vert \coloneqq \vert \mathrm{det}(G)\vert$, and the corresponding $U(1)$-gauge transformations for the system are given by
\begin{align}
\begin{aligned}
    A_{M} &\mapsto A_{M}' = A_{M} + \partial_{M}\Lambda\ ,\\
    \Phi &\mapsto \Phi' = \Phi - \partial_{t}\Lambda\ ,\\
    \psi &\mapsto \psi' = \psi\, \ee^{\frac{\ii Q}{\hbar}  \Lambda}\ .
\end{aligned}
\end{align}
With the gauge condition \citep{Ferrari2008, Brandt2017}
\begin{align}
    \Lambda(\textbf{q},y,t) = -\int\limits_{0}^{y} A_{3}(\textbf{q},\tilde{y},t)\ \mathrm{d}\tilde{y}\ \Rightarrow\ A_{3}' = 0\ ,
\label{eq:decoup_gauge}
\end{align}
the CPA enables a decoupling of the effective tangent and normal motions. The resulting effective tangent Schrödinger equation explicitly reads
\begin{align}
    \ii\hbar\,\tilde{\partial}_{t}\chi_{\mathrm{t}} = \left[-\frac{\hbar^{2}}{2m}\,\tilde{\nabla}^{\mu}\tilde{\nabla}_{\mu} + V_{\mathrm{geo}}\, \hat{I}\right] \chi_{\mathrm{t}}
\label{eq:SE_eff_tang}
\end{align}
and contains the geometry-induced scalar potential-like field
\begin{align}
    V_{\mathrm{geo}} = -\frac{\hbar^{2}}{2m}\,\left( M^{2} - K \right)\ ,
\end{align}
where $M$ and $K$ are the mean and Gaussian curvatures of $(\mathcal{M},g)$ \citep{Jost2011}, respectively. In our two-dimensional case it is always attractive \citep{Schuster2003}.
\par
Equation \eqref{eq:decoup_gauge} entails on the vector potential that its space-like component is a tangent vector field when the limiting procedure is executed, $\underset{y\rightarrow 0}{\mathrm{lim}}\textbf{A}'\in \Gamma\left( T\mathcal{M}\right)$ \citep{Jost2011}. Note, in this limit the gauge leaves all the other components as well as the wave function invariant,
\begin{align}
    \underset{y\rightarrow 0}{\mathrm{lim}}A'_{\mu} = A_{\mu}\ ,\quad \underset{y\rightarrow 0}{\mathrm{lim}}\Phi' = \Phi\ ,\quad \underset{y\rightarrow 0}{\mathrm{lim}}\chi_{\mathrm{t}}' = \chi_{\mathrm{t}}\ .
\label{eq:gauge_outcome}
\end{align}
The tangent vector potential is given as the projection of the external field onto the local frames. Exploiting the block-diagonal structure of the metric tensor field of the Riemannian manifold $\left( \mathbb{R}^{3}, G \right)$ obtained from the parametrization $\mathcal{X}$,
\begin{align}
    (G_{M N}) = \begin{pmatrix}
        G_{12} & G_{12} & 0 \\
        G_{21} & G_{22} & 0 \\
        0      & 0      & 1
    \end{pmatrix} = \begin{pmatrix}
        (G_{\mu\nu}) & (0) \\
        (0) & (I_{1})
    \end{pmatrix} \ ,
\end{align}
we write
\begin{widetext}
\begin{align}
   \mathrm{div}\textbf{A}' = \frac{1}{\sqrt{\vert G\vert}}\,\partial_{M}\left( \sqrt{\vert G\vert}\,A'^{M} \right) = \frac{1}{\sqrt{\vert G\vert}}\,\partial_{\mu}\left( \sqrt{\vert G\vert}G^{\mu\nu}\,(A_{\nu} + \partial_{\nu}\Lambda)  \right) +  \frac{1}{\sqrt{\vert G\vert}}\,\partial_{3}\left( \sqrt{\vert G\vert}\,(A_{3} + \partial_{3}\Lambda)  \right)
\end{align}
Due to the gauge condition \eqref{eq:decoup_gauge}, the second term vanishes. On the manifold $(\mathcal{M}, g)$ we find
\begin{align}
    \mathrm{div}_{\mathcal{M}}(\textbf{A}') \coloneqq \underset{y\rightarrow 0}{\mathrm{lim}}\mathrm{div}(\textbf{A}') = \frac{1}{\sqrt{\vert g\vert}}\partial_{\mu}\,\left( \sqrt{\vert g\vert} A^{\mu} \right)
\end{align}
which means a generally nonvanishing effectively induced covariant divergence of the vector potential field that is present for the observer confined to the Riemannian manifold.
\par
In absence of free charges $Q_{\mathrm{f}}$ ($\rho_{\mathrm{f}}$ is the free charge density) we write
\begin{align}
    Q_{\mathrm{f}} &= \varepsilon_{0}\,\int\limits_{\mathbb{R}^{3}} \mathrm{div}(\textbf{E})\,\mathrm{d}^{3}\tilde{x} = -\varepsilon_{0}\,\int\limits_{\mathbb{R}^{3}} \left[ \mathrm{div}(\boldsymbol{\nabla}\Phi) + \partial_{t}\,\mathrm{div}(\textbf{A}) \right]\,\mathrm{d}^{3}\tilde{x} = \int\limits_{\mathbb{R}^{3}} \rho_{\mathrm{f}}\ \mathrm{d}^{3}\tilde{x} = 0 
\end{align}
where $\Phi$ is the scalar potential. As the condition for $Q_{\mathrm{f}}$ is valid in any gauge, we can insert the one from the above construction. Using the  parametrization $\mathcal{X}$, we find for the pull-back into the parameter space
\begin{align}
\begin{aligned}
    Q_{\mathrm{f}} &= -\varepsilon_{0}\,\iint\limits_{\mathcal{Q}_{\mathrm{t}}\times\mathcal{Q}_{\mathrm{n}}} \mathcal{X}^{*} \left[ \mathrm{div}(\boldsymbol{\nabla}\Phi') + \partial_{t}\,\mathrm{div}(\textbf{A}') \right]\,\sqrt{\vert G\vert}\,\mathrm{d}^{2}\tilde{q}\wedge\mathrm{d}\tilde{y} = 0\ . \\
\end{aligned}
\end{align}
Thus, electric charge conservation requires the vanishing of the spatial integral over the divergences resulting from the gauge. Conderning the CPA, we are mainly interested in a thin layer around $\mathcal{M}$. Evaluating the limiting case $y \rightarrow 0$ by means of a perturbative expansion of $Q_{\mathrm{f}}$ in the normal displacement, we obtain (compare \eqref{eq:gauge_outcome})
\begin{align}
\begin{aligned}
     Q_{\mathrm{f}} = -\varepsilon_{0}\,\int\limits_{\mathcal{Q}_{\mathrm{t}}} \left[ \Delta\Phi + \partial_{t}\,\mathrm{div}_{\mathcal{M}}(\textbf{A})\right]\,\sqrt{\vert g\vert}\,\mathrm{d}^{2}\tilde{q} = -\varepsilon_{0}\,\int\limits_{\mathcal{M}} \left[ \Delta \Phi + \partial_{t}\,\mathrm{div}_{\mathcal{M}}(\textbf{A})\right]\,\mathrm{vol}_{g}  = 0
\end{aligned}
\end{align}
as the leading contribution valid on the manifold. If the external electromagnetic field simultaneously fulfills the Lorenz and Coulomb gauges in the hyperspace, it follows $\Phi = C$ with $C\in\mathbb{R}$ a constant. Hence, $\Delta \Phi = 0$ and the integrated effective induced covariant divergence of the vector potential is conserved. But as this term is generically time-dependent with a factorizable time-function it follows that
\begin{align}
    -\varepsilon_{0}\,\int\limits_{\mathcal{M}} \mathrm{div}_{M}(\textbf{A})\ \mathrm{vol}_{g} = 0\ \ \forall\ t\in\mathbb{R}
    \label{eq:induced_charge}
\end{align}
holds.
\end{widetext}
\par
Note, the condition \eqref{eq:decoup_gauge} is not gauge-fixing so that the effective tangent Schrödinger equation \eqref{eq:SE_eff_tang} is still invariant under tangent gauge transformations \citep{Brandt2018}
\begin{align}
    A_{\mu}(\textbf{q}, t) \mapsto A_{\mu}''(\textbf{q}, t) + \partial_{\mu}\tilde{\Lambda}(\textbf{q}, t)\ .
\end{align}
It will become clear in the next section that the formalism would benefit from the possibility to apply a tangent gauge such that $\mathrm{div}_{\mathcal{M}}(\textbf{A}') = 0$. However, modifying this space-dependence would only yield a contribution to the scalar potential for general nontrivial electromagnetic fields, which is again disadvantageous. Therefore, we stay in the above presented gauge and proceed in dealing with the effective covariant divergence.

\section{The KH-Unitary Transformation on the Embedded Manifold} \label{sec:kh_unitary_transformation}
\subsection{Ansatz of the Space- and Time-Dependent Transformation}
The effective tangent Schrödinger equation for a particle minimally coupled to an electromagnetic field is
\begin{align}
    \ii\hbar\,\partial_{t} \chi_{\mathrm{t}}(\textbf{q},t) = \left[ \hat{H}_{0} + \hat{H}_{\mathrm{int}}(\textbf{q},t) \right]\,\chi_{\mathrm{t}}(\textbf{q},t)
\end{align}
where $\hat{H}_{0}$ describes stationary kinetic and potential terms and $\hat{H}_{\mathrm{int}}$ incorporates the space- and time-dependent interaction terms. Formally, the Kramers-Henneberger transformation is a time-dependent unitary transformation given as a smooth path (parameterized by a time variable) in the Hilbert group of the Hilbert space $U(\mathcal{H})$. Our goal is the appliance of the KH-transformation to a Schrödinger particle bound by the effective geometry-induced potential on a Riemannian manifold. Therefore, it is necessary to account for the spatial variation in $\hat{H}_{\mathrm{int}}$  which is introduced by both the metric tensor field and the local representation of the vector potential field. Extending the mapping we write
\begin{align}
\begin{aligned}
    \hat{\Omega}: \mathcal{Q} \times \mathbb{R} &\rightarrow U(\mathcal{H}) \subset \mathfrak{gl}(\mathcal{H})\ ,\\
    (\textbf{q}, t) &\mapsto \hat{\Omega}(\textbf{q}, t) \coloneqq \mathcal{T}\ee^{\ii\hat{S}(\textbf{q}, t)}
\end{aligned}
\end{align}
with the time-ordering meta-operator $\mathcal{T}$ and the normalized action operator
\begin{align}
    \hat{S}(\textbf{q}, t) \coloneqq \frac{1}{\hbar} \int\limits_{-\infty}^{t} \hat{H}_{\mathrm{int}}(\textbf{q}, \tau)\ \mathrm{d}\tau\ .
\label{eq:action_op}
\end{align}
$\mathfrak{gl}(\mathcal{H})$ means the Lie algebra of endomorphisms of the Hilbert space. We consider the effective tangent Schrödinger equation \eqref{eq:SE_eff_tang}. Expanding the covariant derivative, we  specify the KH-transformation action operator as
\begin{widetext}
\begin{align}
\begin{aligned}
    \hat{S}(\textbf{q}, t) = \frac{1}{\hbar}\,\int\limits_{-\infty}^{t}\left[ \frac{\ii\hbar Q}{2m}\,\frac{1}{\sqrt{\vert g\vert}}\partial_{\mu}\left( \sqrt{\vert g\vert} A^{\mu} \right)\hat{I} + \frac{\ii\hbar Q}{m}A^{\mu}\partial_{\mu} + \frac{Q^{2}}{2m}A^{\mu}A_{\mu}\hat{I} + Q\Phi\,\hat{I} \right]\ \mathrm{d}\tau\ .
\label{eq:KH_utrafo_full}
\end{aligned}
\end{align}
\end{widetext}
The nontrivial geometry of the underlying space has an influence through the metric tensor and vector potential fields, which is the reason for the additional terms that will appear in the treatment below. This means that on a manifold, there is an individual time-dependent unitary transformation assigned to every point in space.
\par
Expression \eqref{eq:KH_utrafo_full} in its entirety should be evaluated for an exact solution of the problem under consideration. However, under certain circumstances that we will assume in the following, it can be simplified. Firstly, due to our assumptions on the gauges discussed above (compare \eqref{eq:gauge_outcome} -- \eqref{eq:induced_charge}) the scalar potential term can be dropped. Secondly, for moderate field intensities, we may neglect all terms that are of second order in the vector potential. Defining the the tangent displacement vector field \citep{Henneberger1968}
\begin{align}
    \alpha^{\mu}(\textbf{q},t) \coloneqq -\frac{Q}{m}\,\int\limits_{-\infty}^{t} A^{\mu}(\textbf{q},\tau)\ \mathrm{d}\tau\ ,
\end{align}
the remaining action operator Within these limits is inferred as
\begin{align}
\begin{aligned}
    \hat{S} &= \frac{\ii Q}{m}\,\int\limits_{-\infty}^{t} \left[ \frac{1}{2}\,\mathrm{div}_{\mathcal{M}}(\textbf{A})\hat{I} + A^{\mu}\partial_{\mu} \right]\ \mathrm{d}\tau \\
    &= -\ii\,\left( \frac{1}{2}\,\mathrm{div}_{\mathcal{M}}(\boldsymbol{\alpha})\hat{I} + \alpha^{\mu}\partial_{\mu} \right) \\
    &\doteq -\ii\,\left( \frac{1}{2}\,\mathrm{div}_{\mathcal{M}}(\boldsymbol{\alpha})\hat{I} + \boldsymbol{\nabla}_{\boldsymbol{\alpha}} \right) \\
    &= \frac{1}{\hbar}\,\hat{\textbf{p}}_{\boldsymbol{\alpha}}\ .
\end{aligned}
\end{align}
$\boldsymbol{\nabla}_{\boldsymbol{\alpha}} = \mathcal{L}_{\boldsymbol{\alpha}} = \alpha^{\mu}\boldsymbol{\partial}_{\mu}$ denotes the covariant derivative of a scalar function which  corresponds to the Lie derivative, i.\,e. the rate of change along the flow of the displacement field $\boldsymbol{\alpha}\in\Gamma(T\mathcal{M})$. Furthermore,  because of \eqref{eq:induced_charge}
\begin{align}
    \int\limits_{\mathcal{M}} \mathrm{div}_{M}(\boldsymbol{\alpha})\ \mathrm{vol}_{g} = 0\ \ \forall\ t\in\mathbb{R}
\end{align}
holds for the displacement field obtained from projection. Hence, the hermiticity of the generalized momentum operator
\begin{align}
\begin{aligned}
    \hat{\textbf{p}}_{\boldsymbol{\alpha}} &= -\ii\hbar\,\left\{ \alpha^{\mu}, \boldsymbol{\nabla}_{\mu} \right\} \\
    &=  -\ii\hbar \left( \frac{1}{2}\,\mathrm{div}_{\mathcal{M}}(\boldsymbol{\alpha})\hat{I} + \boldsymbol{\nabla}_{\boldsymbol{\alpha}} \right)\ .
\end{aligned}
\end{align}
is secured \citep{Schuermann2022}. Consequently, $\hat{S}^{\dagger} = \hat{S}$ still holds true after the above approximations and the generated time-dependent transformation is indeed unitary, $\hat{\Omega}\in U(\mathcal{H})$, which implies $\hat{\Omega}^{\dagger}(\textbf{q},t) = \hat{\Omega}^{-1}(\textbf{q},t)\ \forall\ (\textbf{q},t)\in \mathcal{Q}_{\mathrm{t}}\times\mathbb{R}$.
\par
We use $\hat{\Omega}$ to execute the set of KH-unitary transformations of the Schrödinger operator according to
\begin{align}
    \hat{\Omega}\left[ \hat{H}(t) - \ii\hbar\partial_{t} \right]\hat{\Omega}\hat{\Omega}^{\dagger}\chi_{\mathrm{t}} = 0\ ,
\end{align}
and generate the description of the laser-dressed KH-system as the solution $\chi_{\mathrm{t}}^{\mathrm{KH}}(\textbf{q},t) \coloneqq \hat{\Omega}(\textbf{q},t)\chi_{\mathrm{t}}(\textbf{q},t)$ of the resulting equation. Considering the transformations of the Schrödinger operator, the derivative of an exponential operator with respect to some parameter $\lambda\in\mathbb{R}$ is given \citep{Wilcox1967} by
\begin{align}
    \frac{\partial}{\partial \lambda}\ee^{-\beta\hat{H}} = -\int\limits_{0}^{\beta} \ee^{-(\beta - u)\hat{H}} \frac{\partial \hat{H}}{\partial\lambda} \ee^{-u\hat{H}}\ \mathrm{d}u\ .
\label{eq:exp_deriv}
\end{align}
From the definition of the normalized action operator \eqref{eq:action_op} follows that the time-derivative of $\hat{\Omega}$ cancels the time-dependent interaction part so that the effective tangent KH-Schrödinger equation reads
\begin{align}
    \ii\hbar\,\partial_{t}\chi_{\mathrm{t}}^{\mathrm{KH}}(\textbf{q},t) = \hat{H}^{\mathrm{KH}}(\textbf{q},t)\chi_{\mathrm{t}}^{\mathrm{KH}}(\textbf{q},t)\ ,
\end{align}
with the KH-Hamilton operator
\begin{align}
    \hat{H}^{\mathrm{KH}} = -\frac{\hbar^{2}}{2m}\hat{\Omega}\Delta\hat{\Omega}^{\dagger} + \hat{\Omega}V_{\mathrm{geo}}\,\hat{I}\hat{\Omega}^{\dagger}\ .
    \label{eq:kh_hamilton_operator}
\end{align}
This unitary transformation corresponds to a shift to an accelerated frame of reference connected with the laser field \citep{Henneberger1968}. Assuming a time-harmonic linearly polarized vector potential field of the form $A^{\mu}(\textbf{q},t) = A^{\mu}_{0}(\textbf{q}) \cos{(\omega t)}$, the space- and time-dependencies can be separated and $\hat{\Omega}$ is naturally time-ordered. Following the established scheme \citep{Henneberger1968}, both the transformed Laplace-Beltrami operator and geometric (binding) potential-like term are expressible as series expansions that are ordered on behalf of harmonic functions. Their respective contributions to an effective time-independent potential in the KH-system can be identified, while the remaining terms can be viewed as perturbations. For a sophisticated presentation of the model in the framework of atomic physics, we refer the reader to \citep{Henneberger1968, Vivirito1995, Popov1996, Su1991}. We emphasize that the transformation of the Laplace-Beltrami operator is a novelty due to the spatial confinement. Generally, as the KH-transformation is additive, we can treat the terms of the right hand side of \eqref{eq:kh_hamilton_operator} independently.
\subsection{Consideration of the Scalar Geometry-Induced Potential} \label{sec:main_transformation_potential}
We start with the second term which means the transformation of the binding potential. Denoting recursively nested commutators for $n\in\mathbb{N}_{0}$ as
\begin{align}
\begin{aligned}
    \left[ (\hat{A})^{0}, \hat{B} \right] &\coloneqq \hat{B}\ ,\\
    \left[ (\hat{A})^{n}, \hat{B} \right] &\coloneqq \underbrace{\left[ \hat{A},\cdots, [\hat{A}, [\hat{A}\right.}_{n\,\text{times}}\left., \hat{B}]] \cdots \right]\ ,
    \label{eq:def_nested_commutators}
\end{aligned}
\end{align}
and formally applying \footnote{The Lie algebra of endomorphisms $\mathfrak{gl}(\mathcal{H})$ contains (probably) unbounded operators so that the convergence of the operator exponentials and hence the existence of a corresponding Lie group are not guaranteed. Furthermore, the bounded operators would be a necessary prerequisite for the BCH-formula and the Hadamard lemma to secure their convergence. Therefore, additional assumptions concerning the operator domain may be necessary to fully rely on these expansions. Nevertheless, either we can compute all commutators explicitly in this work or only apply the formulas to find approximations.}, an expansion formula for the similarity transformation of a linear operator $\hat{B}$ based on a Lie group \citep{Wilcox1967},
\begin{align}
    \ee^{\hat{A}}\hat{B}\ee^{-\hat{A}} &= \sum\limits_{n=0}^{\infty} \frac{1}{n!}\left[ (\hat{A})^{n}, \hat{B} \right]\ ,
\label{eq:nested_comm_expan}
\end{align}
we compute the transformation by considering the terms $\left[ \left( \frac{\ii}{\hbar} \hat{\textbf{p}}_{\boldsymbol{\alpha}}\right)^{n} ,V_{\mathrm{geo}}\hat{I} \right]$. Details on the calculations are provided in appendix \ref{sec:transformation_potential}. It is found that
\begin{align}
    \hat{\Omega}V_{\mathrm{geo}}\hat{I}\hat{\Omega}^{\dagger} = \ee^{\alpha^{\mu}\partial_{\mu}}V_{\mathrm{geo}}\hat{I}\ee^{- \alpha^{\mu}\partial_{\mu}}\ ,
    \label{eq:transformed_potential}
\end{align}
meaning that the result is independent of the divergence part. Geometrically, this means that the potential field is transported along the flow of the displacement field, as the operator $\exp{\alpha^{\mu}\partial_{\mu}}$ refers to the functional representation thereof \citep{Dubrovin1992}. Concerning local coordinates, this corresponds to $\hat{\Omega}$ acting as a shift operator on the parameter space $\mathcal{Q}_{\mathrm{t}}\subset \mathbb{R}^{2}$ (with the argument $\boldsymbol{\alpha} = \alpha^{\mu} \boldsymbol{\partial}_{\mu}$).
\par
However, successive translations along the coordinate axes in $\mathcal{Q}_{\mathrm{t}}$ require the operator $\mathrm{exp}\{\,\alpha^{\mu} \partial_{\mu}\}$ to be factorizable. For any pair of sufficiently small elements of a Lie algebra, $\hat{X}$ and $\hat{Y}$, the Baker-Campbell-Hausdorff (BCH) formula \citep{Wilcox1967, Hall2015} expresses the solution $\hat{Z}$ within the Lie group multiplication
\begin{align}
    \ee^{\hat{X}} \ee^{\hat{Y}} = \ee^{\hat{Z}}
\end{align}
as an expansion in terms of nested Lie brackets. The expression can be constructed as a series which can be cast as 
\begin{widetext}
\begin{align}
    \hat{Z}(\hat{X},\hat{Y}) = \hat{X} + \hat{Y} + \frac{1}{2}\,\left[ \hat{X}, \hat{Y} \right] + \frac{1}{12}\,\left( \left[ (\hat{X})^{2},\hat{Y}\right] + \left[ (\hat{Y})^{2},\hat{X}\right] \right) \pm \ldots\ .
\label{eq:BCH}
\end{align}
\end{widetext}
Applying the series expansion formally and assuming the validity of the theorem of Schwarz, it follows that the condition enabling the successive transformations reads
\begin{align}
\begin{aligned}
    \mathcal{L}_{\alpha^{1}\boldsymbol{\partial}_{1}}\alpha^{2}\boldsymbol{\partial}_{2} &= \left[ \alpha^{1}\boldsymbol{\partial}_{1}, \alpha^{2}\boldsymbol{\partial}_{2} \right] \\
    &= \alpha^{1}\alpha^{2}_{\ ,1}\boldsymbol{\partial}_{2} - \alpha^{2}\alpha^{1}_{\ ,2}\boldsymbol{\partial}_{1} \overset{!}{=} \hat{\textbf{0}}\ ,
\end{aligned}
    \label{eq:condition_separation}
\end{align}
which is implied by either one of the following sufficient conditions:
\begin{itemize}
    \item[] (I) Each component depends only on one parameter so that $\forall\ \mu\in\{ 1;2 \}: \alpha^{\mu} = \alpha^{\mu}(q^{\mu})$.
    \item[] (II) At least one component vanishes so that $\exists\ \mu\in\{ 1;2 \}: \alpha^{\mu}=0$.
\end{itemize}
Geometrically, as $\mathcal{L}_{\textbf{u}}\textbf{v} = \left[ \textbf{u}, \textbf{v} \right] = \boldsymbol\nabla_{\textbf{u}}\textbf{v} - \boldsymbol{\nabla}_{\textbf{v}}\textbf{u}$ for arbitrary tangent vector fields $\textbf{u},\textbf{v}\in\Gamma(T\mathcal{M})$, condition \eqref{eq:condition_separation} means that the quadriliteral curves given by the projections of the tangent vector fields into the manifold, here given by the two local components of the displacement field, have to be closed \citep{Misner2017}.
\par
Furthermore, $\boldsymbol{\alpha}$ is set by the parameters of the laser field acting on an electronic systems. For a curved nanoscale structure, the geometric potential can act as the source of binding with a binding energy typically on the scale of hundreds of $\mathrm{meV}$, $\vert V_{\mathrm{geo}}\vert \lesssim \mathcal{O}(10^{2}\,\mathrm{meV})$. The laser wavelength of $\lambda_{\mathrm{EM}}\approx \mathcal{O}(10\,\si{\micro\meter})$ sets the minimal focal spot size which is three orders of magnitude larger than the localized orbitals. Thus, the laser field acts homogeneously and retardation effects are negligible. $\boldsymbol{\alpha}$ then depends mainly on the polarization and the propagation direction of the light, and its spatial inhomogeneity is due to the shape of the manifold. Hence we conclude that a configuration where the vector potential is oriented along one coordinate curve and varies along the other is hard to realize. This means, one may declare condition (I) to be valid and situations exclusively related to condition (II) to be less relevant.
\par
By doing so the transformed potential is found to be (see appendices \ref{sec:shift_operator} and \ref{sec:transformation_potential})
\begin{equation} 
    \hat{\Omega}V_{\mathrm{geo}}\,\hat{I}\hat{\Omega}^{\dagger} = F_{\mathrm{geo}} \left(h^{1} + \tau, h^{2} + \tau\right) \hat{I} 
        \eqqcolon F_{\mathrm{geo}} (\vb{h} + \vb*{\tau})\, \hat{I}\ ,
\label{eq:definition_fgeo}
\end{equation}
where $\tau = \alpha_0 \cdot \sin{(\omega t)}$, $F_{\mathrm{geo}}$, as well as $h^1$ and $h^2$ are discussed in appendix \ref{sec:shift_operator}. Note,  the translation $\vb*{\tau}$ only depends on the time while the space-dependence of the vector potential is incorporated by introducing the new set of variables according to which the geometric potential is adapted.
\par
The effective time-independent potential can be found in a way similar to the original one \citep{Henneberger1968}. For this purpose, and as the (transformed) geometric potential is a function on the parameter space $\mathcal{Q}_{\mathrm{t}} \subset \mathbb{R}^{2}$, we refer to the definition
\begin{align}
    \begin{aligned}
   \mathcal{F}(f)(\textbf{k}) &= \frac{1}{(2\pi)^{\frac{n}{2}}}\,\int\limits_{\mathbb{R}^{n}} \ee^{-\ii \textbf{k}\cdot\textbf{x}}f(\textbf{x})\ \mathrm{d}^{n}x\ ,  \\
    \mathcal{F}^{-1}(\hat{f})(\textbf{x}) &= \frac{1}{(2\pi)^{\frac{n}{2}}}\,\int\limits_{\mathbb{R}^{n}} \ee^{\ii \textbf{k}\cdot\textbf{x}} \hat{f}(\textbf{k})\ \mathrm{d}^{n}k\ ,
    \end{aligned}
\label{eq:FT_def}
\end{align}
and express the transformed geometric potential $F_{\mathrm{geo}}$ via its inverse Fourier-transformation ($\mathcal{K}_{\mathrm{t}}\subset \mathbb{R}^{2}$ being the reciprocal parameter space)
\begin{align}
    F_{\mathrm{geo}} \left( \vb{h} + \vb*{\tau} \right) = \frac{1}{2\pi} \int\limits_{\mathcal{K}_{\mathrm{t}}} \ee^{\ii \vb{k} \cdot \left( \vb{h} + \vb*{\tau} \right)} \mathcal{F}\left( F_{\mathrm{geo}}\right) \left( \vb{k} \right) \ \mathrm{d}^2 k\ .
\label{eq:FT_start}
\end{align}
However, the definition \eqref{eq:FT_def} does not yield the correct Fourier transformation on the Riemannian manifold $\mathcal{M}$ as plane waves in the parameter space, $\exp{\ii \textbf{k}\cdot \textbf{x}}$, do not represent the eigenfunctions of the Laplace-Beltrami operator, in general. For our purpose, nevertheless, any basis of its span suffices. For convenience, we thus proceed with the plane wave expansion (for more on this issue we refer to \citep{Helgason2008}).
\par
Due to the harmonic excitation, we can make use of the Jacobi-Anger expansion \citep{NIST2010},
\begin{align}
    \ee^{\ii z\,\mathrm{sin}(\omega t)} = \sum\limits_{n=-\infty}^{+\infty} J_{n}(z)\,\ee^{\ii n\omega t}\ ,\ z\in\mathbb{C}\ ,
\end{align}
and Bessel's integral of the first kind \citep{NIST2010},
\begin{align}
    J_{n}(z) = \frac{1}{2\pi}\,\int\limits_{-\pi}^{\pi} \ee^{\ii\left( n\vartheta - z\,\sin(\vartheta) \right)}\ \mathrm{d}\vartheta\ ,\ n\in\mathbb{Z}\ ,
\end{align}
to obtain
\begin{widetext}
\begin{align}
    F_{\mathrm{geo}} \left( \vb{h} + \vb*{\tau} \right) = \frac{1}{4 \pi^2} \sum\limits_{n = - \infty}^{\infty} \ee^{\ii n \omega t} \int\limits_{\mathcal{K}_{\mathrm{t}}} \ee^{\ii \vb{k} \cdot \vb{h}}\ \mathcal{F}(F_{\mathrm{geo}}) (\vb{k})\ \int\limits_{-\pi}^{\pi} \ee^{\ii (n \vartheta - \vb*{\alpha_0} \cdot \vb{k} \sin(\vartheta))}\ \mathrm{d}\vartheta\ \mathrm{d}^2 k\ .
\end{align}
By executing the inverse Fourier-transformation and reordering utilizing trigonometric identities we find
\begin{align}
    F_{\mathrm{geo}} \left( \vb{h} + \vb*{\tau} \right) = \frac{1}{2 \pi} \int\limits_{-\pi}^{\pi} F_{\mathrm{geo}} (\vb{h} - \vb*{\alpha_0} \sin(\vartheta))\ \mathrm{d} \vartheta + \sum\limits_{n = 1}^{\infty} \frac{1}{\pi} \int\limits_{-\pi}^{\pi} F_{\mathrm{geo}} (\vb{h} - \vb*{\alpha_0} \sin(\vartheta)) \cdot \cos(n \vartheta + n \omega t)\ \mathrm{d} \vartheta\ .
\end{align}
The space- and time-contributions in the integral can be separated so that we are eventually left with
\begin{align}
    F_{\mathrm{geo}} \left( \vb{h} + \vb*{\tau} \right) = F_0 + \sum\limits_{n = 1}^{\infty} \bigl[ F_{n}^{\mathrm{c}} \cos(n \omega t) - F_{n}^{\mathrm{s}} \sin(n \omega t) \bigl] = F_0 + \delta F (t)\ ,
\end{align}
\end{widetext}
where the abbreviations $\mathrm{s}$ and $\mathrm{c}$ indicate sine and cosine, respectively, and the coefficients for $n\in\mathbb{N}$ are given by
\begin{align}
\begin{aligned}
    F_{0} &\coloneqq \frac{1}{\pi} \int\limits_{-\pi}^{\pi} F_{\mathrm{geo}} (\vb{h} - \vb*{\alpha_0} \sin(\vartheta))\ \mathrm{d} \vartheta\ ,\\
    F_{n}^{\mathrm{c}} &\coloneqq \frac{1}{\pi} \int\limits_{-\pi}^{\pi} F_{\mathrm{geo}} (\vb{h} - \vb*{\alpha_0} \sin(\vartheta)) \cos(n \vartheta)\ \mathrm{d} \vartheta\ ,\\
    F_{n}^{\mathrm{s}} &\coloneqq \frac{1}{\pi} \int\limits_{-\pi}^{\pi} F_{\mathrm{geo}} (\vb{h} - \vb*{\alpha_0} \sin(\vartheta)) \sin(n \vartheta)\ \mathrm{d} \vartheta\ .
\end{aligned}
\label{eq:FT_end}
\end{align}
$F_0$ is the generalization of the well-known effective time-independent potential of a charged particle in a strong electromagnetic field. In terms of $V_{\mathrm{geo}}$ it is given by
\begin{align}
    F_{0} = \frac{1}{\pi} \int\limits_{-\pi}^{\pi} V_{\mathrm{geo}} \Bigl(\vb{h}^{-1} \bigl(\vb{h} (\vb{q}) - \vb*{\alpha_0} \sin{(\vartheta)}\bigl)\Bigl)\ \mathrm{d}\vartheta\ .
\end{align}
The remaining terms can be viewed as harmonic perturbations $\delta F (t)$.
\par
We remark that the above Fourier-expansion is the reason why we restricted ourselves to the consideration of noncompact manifolds. However, in cases where $\mathcal{M}$ contains compact degrees of freedom, the treatment does also work. The integral formulas  have to be replaced by discrete sums.

\subsection{Consideration of the Laplace-Beltrami Operator}
The nontrivial geometry of the underlying space produces a space-dependence of the Laplace-Beltrami operator $\Delta$ and the KH-transformation operator $\hat{\Omega}$ with, generally, nonvanishing mutual commutator. Considering equation \eqref{eq:nested_comm_expan} we expect additional terms in the KH-Hamilton operator that mean a generalization of the concept of laser-dressed systems caused by nontrivial spacetime geometries. A closed analytical expression of the transformed operator has not been obtained, and we need to resort to approximations.
\par
Due to its additive structure, we can decompose $\hat{S} \doteq - \ii\, (\frac{1}{2}\,\mathrm{div}_{\mathcal{M}}(\vb*{\alpha})\hat{I} + \alpha^{\mu}\partial_{\mu})$, and derive
\begin{widetext}
\begin{align}
\begin{aligned}
    %
    \left[ \alpha^{\mu}\partial_{\mu}, \frac{1}{2}\,\mathrm{div}_{\mathcal{M}}(\vb*{\alpha})\hat{I} \right] &= \frac{\alpha_0^2}{2} \, \left[ \tilde{\alpha}^{\mu}\partial_{\mu}, \mathrm{div}_{\mathcal{M}} (\tilde{\vb*{\alpha}}) \hat{I} \right]\cdot \sin^2{(\omega t)} \\    
    &= \frac{\alpha_0^2}{2} \, \tilde{\alpha}^{\mu} \left(\mathrm{div}_{\mathcal{M}} (\tilde{\vb*{\alpha}})\right)_{,\mu} \hat{I} \cdot \sin^2{(\omega t)} \\
    &= \frac{\alpha_0^2}{2}\, \tilde{\alpha}^{\mu}\left(\partial_{\mu}\mathrm{ln}(\mathrm{div}_{\mathcal{M}} (\tilde{\vb*{\alpha}}))\right)\, \mathrm{div}_{\mathcal{M}} (\tilde{\vb*{\alpha}})\hat{I}\cdot \sin^2{(\omega t)} \\
    &= \frac{1}{2}\, s\, \mathrm{div}_{\mathcal{M}} (\vb*{\alpha}) \hat{I} \\[2mm]
    \because\ s(\textbf{q},t) &\coloneqq \alpha^{\mu}(\textbf{q},t)\left(\partial_{\mu}\mathrm{ln}(\mathrm{div}_{\mathcal{M}}(\tilde{\vb*{\alpha}}))\right)
\end{aligned}
\label{eq:comm_summands}
\end{align}
\end{widetext}
where time separability was exploited. Using the adjoint Lie algebra endomorphism
\begin{align}
    \mathrm{ad}_{\hat{X}}: \mathfrak{g} \rightarrow \mathfrak{g}\ ,\ \hat{Y} \mapsto \mathrm{ad}_{\hat{X}}(\hat{Y}) \coloneqq \left[ \hat{X}, \hat{Y} \right]
\end{align}
we can express the BCH formula \eqref{eq:BCH} as \citep{Hall2015}
\begin{align}
    \hat{Z} = \hat{X} + \frac{\mathrm{ad}_{\hat{X}}}{\hat{I}-\ee^{-\mathrm{ad}_{\hat{X}}}}\,\hat{Y} + \mathcal{O}(\hat{Y}^{2})\ .
\label{eq:BCH_special}
\end{align}
$\tilde{\mathcal{O}}(\hat{Y}^{2})$ denotes commutators that contain the element $\hat{Y}$ at least two times. If $[\hat{X}, \hat{Y}] = s\hat{Y}$, with a complex-valued function $s$, we can take advantage of the fact that all these terms vanish. Noting that the commutator between $\alpha^{\mu}\partial_{\mu},\frac{1}{2}\,\mathrm{div}_{\mathcal{M}}(\vb*{\alpha})\hat{I}\in \mathfrak{gl}(\mathcal{H})$ given by \eqref{eq:comm_summands} fulfills this condition, we can formally apply the identity \eqref{eq:BCH_special} and find the expansion
\begin{widetext}
\begin{align}
    \hat{Z} = \alpha^{\mu}\partial_{\mu} + \frac{1}{2}\,\mathrm{div}_{\mathcal{M}}(\vb*{\alpha})\hat{I} + \frac{1}{2}\,\left[ \alpha^{\mu}\partial_{\mu}, \frac{1}{2}\,\mathrm{div}_{\mathcal{M}}(\vb*{\alpha})\hat{I} \right] + \mathcal{O}((\alpha^{\mu}\partial_{\mu})^{2})\ ,
\end{align}
\end{widetext}
wherein the commutator is second-order in the vector potential. We neglect such terms (as applied to \eqref{eq:KH_utrafo_full}) and truncate the solution of $\hat{Z}$ leading to the approximate factorization of the KH-transformation
\begin{align}
\begin{aligned}
    \hat{\Omega} = \ee^{\ii\hat{S}} &= \ee^{(\alpha^{\mu}\partial_{\mu} + \frac{1}{2}\,\mathrm{div}_{\mathcal{M}}(\vb*{\alpha})\hat{I})} \\
    &\simeq \ee^{\alpha^{\mu}\partial_{\mu}} \, \ee^{\frac{1}{2}\,\mathrm{div}_{\mathcal{M}}(\vb*{\alpha})\hat{I}}= \hat{\Omega}_{1}\,\hat{\Omega}_{2}\ .
\end{aligned}
\end{align}
The action of this factorized transformation operator is known: $\hat{\Omega}_{1}$ is identified as the shift operator within the parameter space and $\hat{\Omega}_2$ means a position-dependent similarity transformation that scales the object. However, we note that due to the fact that, in general, $\mathrm{div}_{\mathcal{M}}(\boldsymbol{\alpha}) \neq 0$ the approximated transformation is not unitary anymore. 
\par
To evaluate the transformation of the Laplace-Beltrami operator we consider the action of the operators on the wave function explicitly. That is, we consider
\begin{align}
    \hat{\Omega}\Delta\hat{\Omega}^{\dagger}\chi_{\mathrm{t}}^{\mathrm{KH}} \simeq \hat{\Omega}_{1}\hat{\Omega}_{2}\Delta\hat{\Omega}_{2}^{-1}\hat{\Omega}_{1}^{-1}\chi_{\mathrm{t}}^{\mathrm{KH}}
\end{align}
and identify
\begin{align}
    \hat{\Omega}_{1}^{-1}\chi_{\mathrm{t}}^{\mathrm{KH}}(\textbf{q},t) = F_{\chi} (\vb{h} - \vb*{\tau},t)\ .
\end{align}
The calculations are presented in detail within the appendix section \ref{sec:transformation_laplace}. Eventually, the transformed Laplace-Beltrami operator can also be split into an effective time-independent part and harmonic perturbations according to
\begin{align}
    \hat{\Omega} \Delta \hat{\Omega}^{\dagger} \simeq \Delta_0 + \delta\Delta (t)\ ,
\end{align}
where $\Delta_{0}$ and $\delta\Delta$ are given by \eqref{eq:effective_laplace} and \eqref{eq:harmonic_laplace}, respectively. Finally, the Schrödinger equation describing the tangent laser-dressed system reads
\begin{align}
    \ii\hbar\,\partial_{t}\chi^{\mathrm{KH}}_{\mathrm{t}} = \left[ \Delta_{0} + F_{0}\,\hat{I} + \delta\Delta(t) + \delta F(t)\hat{I} \right]\,\chi^{\mathrm{KH}}_{\mathrm{t}}\ .
\end{align}
This equation is the main result of this article.
\par
The KH-approximation amounts to determining the laser-dressed KH-spectrum based on the time-independent part, when the other terms act perturbatively. This grants access to all usually studied aspects of KH-light matter interaction such as in \citep{Popov1999, Volkova1997, Su1990, Vivirito1995, Popov2003, Morales2011, Yamanouchi2021}. Here, we note important differences to the atomic case \citep{Henneberger1968, Popov1996}. Firstly, the equation represents a dimensionally-reduced system placed on a geometrically nontrivial Riemanninan manifold. Consequently, the kinetic energy operator admits and yields a space-dependent Laplace-Beltrami operator accompanied by a geometry-induced scalar correction term \citep{Jensen1971, Costa1981, Wang2017}. The latter may be understood as the binding potential, which can thus be identified as a pure quantum mechanical effect itself. Furthermore, the nontrivial geometry requires a specific gauge condition for the electromagnetic potentials, resulting in a space-dependent unitary operator that conveys transport along the flow of the displacement vector field $\boldsymbol{\alpha}$. Our considerations lead to a qualitatively different KH-Schrödinger equation which includes modified time-independent differential and potential operators. The same is true for the time-dependent perturbative terms for which we observe the appearance of differential operator-valued terms. We note that the dimensionally-reduced setting is topologically different to the original one. However, the mathematical treatment of the transformation is identical. Thus, this derivation may be viewed as a generalized formalism. If the manifold is flat, the displacement field $\boldsymbol{\alpha}$ will be homogeneous, so the KH-unitary operator reduces to the shift operator; or more precisely even the translation operator.
\par
Mathematically, our new effective tangent KH-Schrödinger equation is different from its three-dimensional relatives. The investigation of its spectrum and eigenmodes are ongoing research.

\section{Conclusion} \label{sec:conclusion}
We performed a KH-unitary transformation for a laser-driven charged particle constrained to move on a two-dimensional Riemannian manifold, embedded in a three-dimensional hyperspace, with non-trivial metric tensor. The geometry-induced potential-like term was considered as the binding potential. We utilized a gauge condition under which an electromagnetic field can act on the confined particles. Because of this, the geometrical properties of the manifold determine an effective spatial structuring of the vector potential. This renders the KH-unitary transformation both space- and time-dependent. We deduced a KH-Schrödinger equation in which the Laplace-Beltrami operator contributes in a nontrivial way. Our results generalize the conventional KH-approach as a first step towards understanding possible laser-dressed states hosted by non-trivial spaces of reduced dimension.

\section*{Acknowledgements} This work has been supported by the Deutsche Forschungsgemeinschaft project number 429194455. We benefited from discussions with Lars Meschede, Alexandra Schrader,  and Dominik Schulz.

\appendix

\section{The Shift Operator}
\label{sec:shift_operator}
In its original version \citep{Henneberger1968}, the KH-unitary transformation is given by a time-dependent translation operator of the form $\mathrm{exp}(\alpha^i \partial_i)$, so that the conditions $\partial_i \alpha^i = 0$ are satisfied for all $i\in\{1;2;3\}$ separately (no implicit summation). However, in our generalized case, this cannot be guaranteed and, consequently, the well-known representation of the translation operator as a Taylor expansion of a shifted function cannot be applied. To analyze the transformation encountered in this work (see \eqref{eq:shift_potential} and \eqref{eq:shift_laplace}), we have to consider the action of the operator $\hat{\Omega} = \mathrm{exp}\left\{\tau \cdot v(x)\, \partial_x\right\}$ on a function $f = f(x)$, where $\tau$ is independent of $x$ and $\partial_x v \neq 0$, in general. A solution of this problem can be constructed using the concepts of function iteration, yielding the shift operator.
\par
Two mappings of a topological space $(X,\mathcal{T})$ to itself, $s,g: X\rightarrow X$, are called topological conjugate if there exists a homeomorphism $h$ such that
\begin{align}
    g = h^{-1}\circ s\circ h\ \Leftrightarrow\ g(x) = h^{-1}(s(h(x)))
\end{align}
for all $x\in X$. Let us denote the translation operator acting on a scalar function $f$ that is dependent on the variable $x\in\mathbb{R}$ as $\hat{T}_{x}[\tau]$. Its effect is the transformation of the argument according to the specific choice $s(x) = x+\tau$, with $\tau\in\mathbb{R}$, such that $\hat{T}_x[\tau]f(x) = f(s(x)) = f(x+\tau)$. The shift operator, correspondingly, implies a topological conjugate transformation $g$. Thus when applied, the conjugation map has to be incorporated as a change of coordinates $x\mapsto h(x)$ which is defined as
\begin{align}
\begin{aligned}
    h(x) &= \int \frac{1}{v(x)}\ \mathrm{d}x\ ,\\
    f(x) &= F(h(x))\ \Leftrightarrow\ f = F \circ h
\end{aligned}
\label{eq:transformation_shift_operator}
\end{align}
for all $x\in X$. With that, the action of the shift operator can be expressed as
\begin{align}
\begin{aligned}
    \hat{\Omega}f(x) &= \ee^{\tau \cdot v(x)\, \partial_x} f(x) = \ee^{\tau\, \partial_{h(x)}} F(h(x)) \\
    &= \hat{T}_{h}[\tau]F(h(x)) = F(h + \tau)\\
    &= f(g_{\tau}(x)) = f(h^{-1}(h(x)+\tau))\ .
\end{aligned}
\label{eq:action_shift_operator}
\end{align}
Observe that $h$ is an Abel function, meaning a solution of the Abel equation based on the topological conjugate of the translation action,
\begin{align}
    h(g(x)) = h(x) + 1\ .
\end{align}
As such, it appears as the canonical coordinate to parametrize Lie advective flows for which the shift operator here is an example. For further details, we refer to \citep{Hamermesh1989, Aczel2006, Hall2015}. The common translation operator is included in this treatment and can be reproduced by choosing $v(x) = 1$, for which $g=s\ \Leftrightarrow\ h=\mathrm{id}_{X}$.
\par
In order to describe two-dimensional Riemannian manifolds, we need to apply this procedure to a multi variable function $f=f(\textbf{x})$. Then, in general $\hat{\Omega} = \mathrm{exp}(\ii\, v^{\mu}(\textbf{x})\,\partial_{\mu})$ but any general expression of the transformation is rather complicated as mutual dependencies of all arguments need to be considered. For our purposes (see \eqref{eq:condition_separation}), it is sufficient to assume two successive shift operators acting as
\begin{align}
    \hat{\Omega}f(\textbf{x}) = \ee^{\tau\cdot v^{2}\partial_{2}}\,\ee^{\tau\cdot v^{1}\partial_{1}}f(\textbf{x}) = f(\textbf{g}_{\tau}(\textbf{x}))\ ,
\end{align}
now with a two-component function $g_{\tau}: X\times X \rightarrow X \times X$. Thus, we define the Abel functions with respect to a certain argument (no summation in $\mu$) but potentially dependent on all of them, as
\begin{align}
    h^{\mu}(\textbf{x}) \coloneqq \int \frac{1}{v^{\mu}(\textbf{x})}\ \mathrm{d}x^{\mu}\ .
\end{align}
As our situation is connected to the validity of either one of two specific cases leading to the quadriliteral closure condition and we only focus on one of them as discussed in sec. \ref{sec:main_transformation_potential}, we will restrict our further considerations to this. Thereby, further simplifications are achieved.
\par
Assume $\forall\ \mu\in\{ 1;2 \}: v^{\mu} = v^{\mu}(x^{\mu})$. Then $h^{\mu} = h^{\mu}(x^{\mu})$, too, and thus the components of $\textbf{g}_{\tau}$ are given by
\begin{align}
    g_{\tau}^{\mu}(x^{\mu}) = \left( (h^{\mu})^{-1}\circ s \circ h^{\mu} \right)(x^{\mu})\ .
\end{align}
This case means that both variables are shifted separately. We can define the transformed function as
\begin{align}
    \hat{\Omega}f(\textbf{x}) = F\left(h^{1}(x^{1}) + \tau, h^{2}(x^{2})+\tau \right)\ .
\end{align}
We verified by mathematical induction that
\begin{align}
    \left(\partial_{h}^{n}F\right)\circ s\circ h = \left[ \frac{1}{(\partial_{x}h)\circ g_{\tau}}\,\partial_{g_{\tau}} \right]^{n}f\circ g_{\tau}
\end{align}
holds for all $n\in\mathbb{N}_{0}$. Thus, we find in particular that
\begin{align}
\begin{aligned}
    F'(h+\tau) &= v(g_{\tau})\,f'(g_{\tau})\ , \\
    F''(h +\tau) &= v(g_{\tau})^{2}\,f''(g_{\tau}) + v(g_{\tau})\,v'(g_{\tau})\,f'(g_{\tau})\ .
    \label{eq:chi_derivatives}
\end{aligned}
\end{align}
We can use this formula and the definition \eqref{eq:transformation_shift_operator} to evaluate the action of the derivatives on the transformed function and find, up to second-order
\begin{align}
\begin{aligned}
    \partial_{x}F(h + \tau) &= \frac{1}{v(x)}\,\partial_{h}F(h+\tau)\ , \\
    \partial_{x}^{2}F(h + \tau) &= \frac{1}{v(x)^{2}}\,\left( \partial_{h}^{2}F(h+\tau) - \partial_{x}v(x)\,\partial_{h}F(h+\tau) \right)\ .
    \label{eq:F_derivatives}
\end{aligned}
\end{align}
If we assume $f$ to be a function that is dependent on two variables $x$ and $y$ which are shifted successively and independently -- i.\,e. $v^x = v^x(x)$ and $v^y = v^y (y)$ and denoting $F(h^x + \tau^x, h^y + \tau^y) = F(\vb{h} + \vb*{\tau})$, $f(g^x_{\tau}, g^y_{\tau}) = f(\vb{g})$ -- then the second order mixed derivative of the shifted function is given by
\begin{align}
\begin{aligned}
    \partial_y \partial_x F(\vb{h} + \vb*{\tau}) &= \frac{1}{v^y v^x} \partial_{h^y} \partial_{h^x} F(\vb{h} + \vb*{\tau}) \\
    \partial_{h^y} \partial_{h^x} F(\vb{h} + \vb*{\tau}) &= v^y (g^y) v^x (g^x) \partial_{g^y} \partial_{g^x} f (\vb{g}_{\tau})\ .
\end{aligned}
\end{align}

\section{Evaluation of the KH-Unitary Transformation}
\subsection{Unitary Transformation of the Geometric Potential} \label{sec:transformation_potential}
Here we consider the space- and time-dependent unitary KH-transformation of the potential leading to \eqref{eq:transformed_potential}. According to \eqref{eq:nested_comm_expan} it can be expressed as
\begin{align}
    \hat{\Omega}V_{\mathrm{geo}}\,\hat{I}\hat{\Omega}^{\dagger} = \sum\limits_{n=0}^{\infty} \frac{1}{n!}\left[ (\ii\hat{S})^{n}, V_{\mathrm{geo}}\,\hat{I} \right]\ .
    \label{eq:transformed_potential_comm}
\end{align}
As the geometric potential is a function of parameter space, we evaluate the commutators with the parameter position operator. It follows
\begin{align}
    \begin{aligned}
        \left[ (\ii\hat{S})^{0}, \hat{q} \right] &= \hat{q} = \left[ (\alpha^{\mu}\partial_{\mu})^0, \hat{q} \right]\ ,\\
        \left[ (\ii\hat{S})^{1}, \hat{q} \right] &= \left[ \frac{1}{2}\,\mathrm{div}_{\mathcal{M}}(\vb*{\alpha})\hat{I} + \alpha^{\mu}\partial_{\mu}, \hat{q} \right] = \left[ \alpha^{\mu}\partial_{\mu}, \hat{q} \right]\ ,
    \end{aligned}
\end{align}
as $\left[ \frac{1}{2}\,\mathrm{div}_{\mathcal{M}}(\vb*{\alpha})\hat{I}, \hat{q} \right] = \hat{0}$. Assuming that the nested commutator term of some order $n\in\mathbb{N}$ in \eqref{eq:transformed_potential_comm} satisfies the equation
\begin{equation}
    \left[ (\ii\hat{S})^{{n}}, \hat{q} \right] = \left[ (\alpha^{\mu}\partial_{\mu})^{n}, \hat{q} \right] \ ,
    \label{eq:induction_base}
\end{equation}
one can show by mathematical induction that the condition holds for all $n\in\mathbb{N}$. As any commutator of a function with the derivative operator simply results in another space-dependent function, it becomes clear that the following term vanishes:
\begin{align}
    \begin{aligned}
        \left[ \frac{1}{2}\,\mathrm{div}_{\mathcal{M}}(\vb*{\alpha})\hat{I}, \left[ (\alpha^{\mu}\partial_{\mu})^{n-1}, \hat{q} \right] \right] &= \hat{0}\ .
            \label{eq:induction_step1}
    \end{aligned}
\end{align}
A combination of the above statements leads to
\begin{align}
    \begin{aligned}
        \left[ (\ii\hat{S})^{n}, \hat{q} \right] &\hspace{0.4mm}\stackrel{\eqref{eq:def_nested_commutators}}{=}\hspace{0.4mm} \left[ \left(\frac{1}{2}\,\mathrm{div}_{\mathcal{M}}(\vb*{\alpha})\hat{I} + \alpha^{\mu}\partial_{\mu}\right), \left[ (\ii\hat{S})^{n-1}, \hat{q} \right] \right] \\
            &\stackrel{\eqref{eq:induction_base}}{=} \left[  \left( \frac{1}{2}\,\mathrm{div}_{\mathcal{M}}(\vb*{\alpha})\hat{I} + \alpha^{\mu}\partial_{\mu} \right), \left[ ( \alpha^{\mu}\partial_{\mu})^{n-1}, \hat{q} \right] \right] \\
            &\stackrel{\eqref{eq:induction_step1}}{=} \left[ \alpha^{\mu}\partial_{\mu}, \left[ (\alpha^{\mu}\partial_{\mu})^{n-1}, \hat{q} \right] \right] \\
            &\hspace{1.5mm}=\hspace{1.5mm} \left[( \alpha^{\mu}\partial_{\mu})^{n}, \hat{q} \right]\ .
    \end{aligned}
\end{align}
Eventually, we obtain the representation
\begin{align}
    \hat{\Omega}\,\hat{q}\,\hat{\Omega}^{\dagger} = \ee^{\alpha^{\mu}\partial_{\mu}} \hat{q}\, \ee^{-\alpha^{\mu}\partial_{\mu}}\ ,
\end{align}
and by analyticity of the potential field
\begin{align}
    \hat{\Omega}V_{\mathrm{geo}}\,\hat{I}\hat{\Omega}^{\dagger} = \ee^{\alpha^{\mu}\partial_{\mu}} V_{\mathrm{geo}}\,\hat{I}\, \ee^{-\alpha^{\mu}\partial_{\mu}}\ ,
    \label{eq:shift_potential}
\end{align}
where $\mathrm{exp}\{\alpha^{\mu}\partial_{\mu}\}$, the functional representation of the flow along $\boldsymbol{\alpha}$ \citep{Dubrovin1992}, acts as a shift operator (see appendix \ref{sec:shift_operator}). As such, the transformation \eqref{eq:shift_potential} can be executed. Specifically, we have $X = \mathcal{Q}_{\mathrm{t}} \subset \mathbb{R}^{2}$ the parameter space, $\textbf{v} = \tilde{\boldsymbol{\alpha}} (\vb{q})$ describing merely the space-dependence of the shift, and $\tau = \alpha_0 \cdot \sin{(\omega t)}$ containing the amplitude of the vector potential and the time-dependence and therefore
\begin{align}
    \vb*{\alpha} (\vb{q}, t) = \vb*{\tilde{\alpha}} (\vb{q}) \cdot\alpha_0 \cdot \sin{(\omega t)}\ .
    \label{eq:alpha_parts}
\end{align}
Then, the result of \eqref{eq:shift_potential} is $F_{\mathrm{geo}} = V_{\mathrm{geo}} (g_{\tau}(\vb{q}))$. It can be treated in the same way as the binding potential in \citep{Henneberger1968}. Further explanations can be found in the main text.

\begin{widetext}
\subsection{Approximate Derivation of the Transformed Laplace-Beltrami Operator} \label{sec:transformation_laplace}
The Laplace-Beltrami operator is a differential operator rather than a function of position, and to our best knowledge no closed form of its commutation with the KH-transformation operator $\hat{\Omega}$ is known. Therefore, we consider an approximate factorization of the latter according to the form
\begin{align}
    \hat{\Omega} \simeq \hat{\Omega}_{1}\,\hat{\Omega}_{2}\ \Rightarrow\ \hat{\Omega}\Delta\hat{\Omega}^{\dagger} \simeq \hat{\Omega}_{1}\,\hat{\Omega}_{2} \Delta \hat{\Omega}_{2}^{-1}\,\hat{\Omega}_{1}^{-1}
    \label{eq:shift_laplace}
\end{align}
that we derived relying on the BCH-formula (details are in the main text). Note that this approximation demands the tribute of unitarity. Indeed, using \eqref{eq:nested_comm_expan} and the formula
\begin{align}
\begin{aligned}
    \Delta(fg) = f\Delta g + 2 g(\boldsymbol{\nabla}f,\boldsymbol{\nabla}g) + g\Delta f = f\,\Delta g + 2g^{\mu\nu}f_{,\mu}\,g_{,\nu} + g\,\Delta f\ ,
\end{aligned}
\end{align}
giving the action of the Laplace-Beltrami operator on a product of scalar functions, it is easy to verify that the inner transformation yields
\begin{align}
\begin{aligned}
    \hat{\Omega}_{2}\Delta\hat{\Omega}_{2}^{-1} &= \Delta - \frac{1}{2} \Delta\left( \mathrm{div}_{\mathcal{M}}(\boldsymbol{\alpha}) \right)\hat{I} - \boldsymbol{\nabla}\mathrm{div}_{\mathcal{M}}(\boldsymbol{\alpha}) + \frac{1}{4} g \left( \boldsymbol{\nabla}\mathrm{div}_{\mathcal{M}}(\boldsymbol{\alpha}), \boldsymbol{\nabla}\mathrm{div}_{\mathcal{M}}(\boldsymbol{\alpha}) \right) \hat{I} \\
        &= \underbrace{g^{\mu\nu} \vphantom{\left[ \frac{1}{\sqrt{g}}\right]}}_{\eqqcolon A^{\mu\nu}} \partial_{\mu}\partial_{\nu} + \underbrace{\left[ \frac{1}{\sqrt{g}} \left( \sqrt{g} g^{\mu\nu} \right)_{,\mu} - g^{\mu\nu} (\mathrm{div}(\vb*{\alpha}))_{,\mu} \right]}_{\eqqcolon B^{\nu}} \partial_{\nu} + \underbrace{\left[ -\frac{1}{2} \Delta (\mathrm{div} (\vb*{\alpha})) + \frac{1}{4} g^{\mu\nu} (\mathrm{div} (\vb*{\alpha}))_{,\mu} (\mathrm{div} (\vb*{\alpha}))_{,\nu} \right]}_{\eqqcolon C}\hat{I} \\
        &= A^{\mu\nu}\, \partial_{\mu}\partial_{\nu} + B^{\mu}\, \partial_{\mu} + C\, \hat{I}\ .
\end{aligned}
\end{align}
Now, the remaining operators can act sequentially on the shifted KH wave function
\begin{align}
    \hat{\Omega}_{1}^{-1} \chi_{\mathrm{t}}^{\mathrm{KH}} (\vb{q}, t) = F_{\chi} (\vb{h} - \vb*{\tau}, t)\,
\end{align}
relying on the action of the shift operator. By making use of the identities \eqref{eq:F_derivatives}, we obtain
\begin{align}
    \hat{\Omega}_{2}\Delta\hat{\Omega}_{2}^{-1} F_{\chi} (\vb{h} - \vb*{\tau}, t) &= \left[ \frac{A^{\mu\nu}}{\tilde{\alpha}^{\mu} \tilde{\alpha}^{\nu}} \partial_{h^{\mu}} \partial_{h^{\nu}}  - \frac{A^{\mu\mu}}{(\tilde{\alpha}^{\mu})^2} \tilde{\alpha}^{\mu}_{,\mu} \partial_{h^{\mu}} + \frac{B^{\nu}}{\tilde{\alpha}^{\nu}} \partial_{h^{\nu}}  + C\,\hat{I} \right] F_{\chi} (\vb{h} - \vb*{\tau}, t)\ .
\end{align}
Note, that we exploited the fact that $\alpha^{\mu} = \alpha^{\mu} (q^{\mu})$. Application of \eqref{eq:chi_derivatives} yields
\begin{align}
\begin{aligned}
    \hat{\Omega}_{2}\Delta\hat{\Omega}_{2}^{-1} F_{\chi} (\vb{h} - \vb*{\tau}, t) = &\left\{ \frac{A^{\mu\nu}}{\tilde{\alpha}^{\mu} \tilde{\alpha}^{\nu}} \left[ \tilde{\alpha}^{\mu} \tilde{\alpha}^{\nu} \circ \vb{g}_{\tau}^{-1} \right] \partial_{g_{\tau}^{\mu}} \partial_{g_{\tau}^{\nu}}  + \left( \frac{A^{\mu\mu}}{\tilde{\alpha}^{\mu} \tilde{\alpha}^{\mu}} \left[ \tilde{\alpha}^{\mu} \tilde{\alpha}^{\mu}_{,\mu} \circ \vb{g}_{\tau}^{-1} \right] \right. \right. \\
        & \left. \left. - \frac{A^{\mu\mu}}{(\tilde{\alpha}^{\mu})^2} \tilde{\alpha}^{\mu}_{,\mu} \left[ \tilde{\alpha}^{\mu} \circ \vb{g}_{\tau}^{-1} \right] + \frac{B^{\mu}}{\tilde{\alpha}^{\mu}} \left[ \tilde{\alpha}^{\mu} \circ \vb{g}_{\tau}^{-1} \right] \right) \partial_{g_{\tau}^{\mu}} + C \right\} \chi_{\mathrm{t}}^{\mathrm{KH}} (\vb{g}_{\tau}^{-1}(\vb{q}), t) \ .
\end{aligned}
\end{align}
After one final application of the shift operator $\hat{\Omega}_1$ the coefficients will be evaluated at the point $\vb{g}_{\tau}(\vb{q}) = \vb{h}^{-1} (\vb{h}(\vb{q}) + \vb*{\tau})$ and the derivatives of the previously shifted wave function transform as follows: $\hat{\Omega}_{1} \partial_{g_{\tau}^{\mu}} \chi_{\mathrm{t}}^{\mathrm{KH}} (\vb{g}_{\tau}^{-1}(\vb{q}), t) = \partial_{\mu} \chi_{\mathrm{t}}^{\mathrm{KH}} (\vb{q}, t)$, $\hat{\Omega}_{1} \partial_{g_{\tau}^{\nu}} \partial_{g_{\tau}^{\mu}} \chi_{\mathrm{t}}^{\mathrm{KH}} (\vb{g}_{\tau}^{-1}(\vb{q}), t) = \partial_{\nu} \partial_{\mu} \chi_{\mathrm{t}}^{\mathrm{KH}} (\vb{q}, t)$ which results in
\begin{align}
\begin{aligned}
    \hat{\Omega}_{1}\hat{\Omega}_{2}\Delta\hat{\Omega}_{2}^{-1} F_{\chi} (\vb{h} - \vb*{\tau}, t) = &\left\{ \left[ \frac{A^{\mu\nu}}{\tilde{\alpha}^{\mu} \tilde{\alpha}^{\nu}} \circ \vb{g}_{\tau} \right] \tilde{\alpha}^{\mu} \tilde{\alpha}^{\nu} \partial_{{\mu}} \partial_{{\nu}}  + \left( \left[\frac{A^{\mu\mu}}{\tilde{\alpha}^{\mu} \tilde{\alpha}^{\mu}} \circ \vb{g}_{\tau} \right] \tilde{\alpha}^{\mu} \partial_{\mu} \tilde{\alpha}^{\mu} \right. \right. \\
        & \left. \left. - \left[ \frac{A^{\mu\mu}}{(\tilde{\alpha}^{\mu})^2} \partial_{\mu} \tilde{\alpha}^{\mu} \circ
        \vb{g}_{\tau} \right] \tilde{\alpha}^{\mu} + \left[ \frac{B^{\mu}}{\tilde{\alpha}^{\mu}} \circ \vb{g}_{\tau} \right] \tilde{\alpha}^{\mu} \right) \partial_{{\mu}} + \left[ C \circ \vb{g}_{\tau} \right] \right\} \chi_{\mathrm{t}}^{\mathrm{KH}} (\vb{q}, t) \ .
    \label{eq:laplace_coeffs}
\end{aligned}
\end{align}
In the following we want to extract the effective time-independent contribution to the Laplace-Beltrami operator. This will be done using a modified version of the sorting scheme already employed for the binding potential. Precisely, we group all coefficient functions of the operators in \eqref{eq:laplace_coeffs} according to their explicit time-dependence. Introducing the notation
\begin{align}
    M_{\vb{m}} = \sum_{i=0}^2 \tilde{M}_{\vb{m}}^{(i)} \alpha_0^i \sin^i{(\omega t)}\ ,
\end{align}
where $\textbf{m}\in\{ 0;1;2 \}^{2}\subset \mathbb{N}_{0}^{2}$ is a multi-index, equation \eqref{eq:laplace_coeffs} takes on the form
\begin{align}
     \hat{\Omega}_{1}\hat{\Omega}_{2}\Delta\hat{\Omega}_{2}^{-1} F_{\chi} (\vb{h} - \vb*{\tau}, t) = M_{\textbf{m}}\partial^{\textbf{m}}
\end{align}
and the term $\propto \mathrm{sin}^{i}(\omega t)$ with $i\in\{ 0;1;2 \}$, which is included in the term $\mathrm{div}(\boldsymbol{\alpha})$, can be extracted. The result thereof is displayed in tab. \ref{tab:laplace_coeffs}. Note that the implicit contribution due to the transformation of the argument based on the function $\vb{g}_{\tau}$ is not relevant for this scheme, so $\tilde{M}_{\textbf{m}}^{(i)}$ can still carry time-dependent terms. Because of this, these coefficients can be treated analogous to the binding potential \eqref{eq:FT_start} - \eqref{eq:FT_end}, which results in their representation by a Fourier series including time-independent integrals combined with sine and cosine contributions (compare \eqref{eq:FT_end}), delivering
\begin{align}
    \tilde{M}_{\vb{m}}^{(i)} = \tilde{M}_{\vb{m}, 0}^{(i)} + \sum_{n=1}^{\infty} \left( \tilde{M}_{\vb{m}, n}^{(i) \mathrm{c}} \cos{(n \omega t)} - \tilde{M}_{\vb{m}, n}^{(i) \mathrm{s}} \sin{(n \omega t)} \right)\ .
\end{align}
\begin{table}
    \caption{Coefficients $\tilde{M}^{(i)}_{\textbf{m}}$ of the operator contributions to the Laplace-Beltrami operator according to their explicit time-dependencies (i.\,e. without $\vb{g}_{\tau}$). The notation $\tilde{\vb*{\alpha}}$ means that only the space dependent contributions to the function $\vb*{\alpha}$ are considered (see \eqref{eq:alpha_parts}).}
        \begin{tabular}{c || >{\centering\arraybackslash}p{0.4\textwidth} | >{\centering\arraybackslash}p{0.25\textwidth} | >{\centering\arraybackslash}p{0.25\textwidth}}
            \toprule[1pt]
            $\tilde{M}^{(i)}_{\textbf{m}}$  &  $\propto \sin^0{(\omega t)}$ &   $\propto \sin^1{(\omega t)}$    &   $\propto \sin^2{(\omega t)}$   \\
            \midrule
            $\partial_{\mu}\partial_{\nu}$  &   $\left[ \frac{g^{\mu\nu}}{\tilde{\alpha}^{\mu} \tilde{\alpha}^{\nu}} \circ \vb{g}_{\tau} \right] \tilde{\alpha}^{\mu} \tilde{\alpha}^{\nu}$   &  --   &  --   \\[0.5cm]
            $\partial_{\mu}$  &  $\left[\frac{g^{\mu\mu}}{\tilde{\alpha}^{\mu} \tilde{\alpha}^{\mu}} \circ \vb{g}_{\tau} \right] \tilde{\alpha}^{\mu} \tilde{\alpha}^{\mu}_{,\mu} - \left[ \frac{g^{\mu\mu}}{(\tilde{\alpha}^{\mu})^2} \tilde{\alpha}^{\mu}_{,\mu} \circ
        \vb{g}_{\tau} \right] \tilde{\alpha}^{\mu} + \left[ \frac{\frac{1}{\sqrt{g}} \left( \sqrt{g} g^{\nu\mu} \right)_{,\nu}}{\tilde{\alpha}^{\mu}} \circ \vb{g}_{\tau} \right] \tilde{\alpha}^{\mu}$  &   $\left[ - \frac{g^{\nu\mu} (\mathrm{div}(\vb*{\tilde{\alpha}}))_{,\nu}}{\tilde{\alpha}^{\mu}} \circ \vb{g}_{\tau} \right] \tilde{\alpha}^{\mu}$    &   --   \\[0.5cm]
            $\hat{I}$  &    --   &   $\left[ -\frac{1}{2} \Delta (\mathrm{div} (\vb*{\tilde{\alpha}})) \circ \vb{g}_{\tau} \right]$  & $\left[ \frac{1}{4} g^{\mu\nu} (\mathrm{div} (\vb*{\tilde{\alpha}}))_{,\mu} (\mathrm{div} (\vb*{\tilde{\alpha}}))_{,\nu} \circ \vb{g}_{\tau} \right]$   \\
            \bottomrule
        \end{tabular}
    \label{tab:laplace_coeffs}
\end{table} \\
Now, the inherent time-dependencies are fully displayed, and the temporal contributions of the Laplace-Beltrami operator can be sorted. Applying trigonometric addition theorems, we find that an effective time-independent term is given by
\begin{align}
    \left( \hat{\Omega} \Delta \hat{\Omega}^{\dagger} \right)_0 = \left[ \Tilde{M}_{\textbf{m},0}^{(0)} - \frac{\alpha_0}{2} \Tilde{M}_{\textbf{m},1}^{(1)\mathrm{s}} + \frac{\alpha_0^2}{2} \left( \Tilde{M}_{\textbf{m},0}^{(2)} - \frac{1}{2} \Tilde{M}_{\textbf{m},2}^{(2)\mathrm{c}} \right) \right] \partial^{\textbf{m}} \ .
    \label{eq:effective_laplace}
\end{align}
In the same manner we find that the harmonic perturbation contributions of the operator follow a general expression for $n \geq 3$, and read
\begin{align}
\begin{aligned}
    \left( \hat{\Omega} \Delta \hat{\Omega}^{\dagger} \right)_1 =\, &\frac{1}{2} \left[ 2 \Tilde{M}_{\textbf{m},1}^{(0)\mathrm{c}} - \alpha_0 \Tilde{M}_{\textbf{m},2}^{(1)\mathrm{s}} + \frac{\alpha_0^2}{2} \left( - \Tilde{M}_{\textbf{m},3}^{(2)\mathrm{c}} + \Tilde{M}_{\textbf{m},1}^{(2)\mathrm{c}} \right) \right] \cos{(\omega t)} \,\partial^{\textbf{m}} \\
        &+ \frac{1}{2} \left[ 2 \Tilde{M}_{\textbf{m},1}^{(0)\mathrm{s}} + \alpha_0 \left( 2 \Tilde{M}_{\textbf{m},0}^{(1)} - \Tilde{M}_{\textbf{m},2}^{(1)\mathrm{c}} \right) + \frac{\alpha_0^2}{2} \left( \Tilde{M}_{\textbf{m},3}^{(2)\mathrm{s}} - 3 \Tilde{M}_{\textbf{m},1}^{(2)\mathrm{s}} \right) \right] \sin{(\omega t)} \, \partial^{\textbf{m}}\ , \\[3mm]
    \left( \hat{\Omega} \Delta \hat{\Omega}^{\dagger} \right)_2 =\, &\frac{1}{2} \left[ 2 \Tilde{M}_{\textbf{m},2}^{(0)\mathrm{c}} + \alpha_0 \left( - \Tilde{M}_{\textbf{m},3}^{(1)\mathrm{s}} + \Tilde{M}_{\textbf{m},1}^{(1)\mathrm{s}} \right) + \alpha_0^2 \left( \Tilde{M}_{\textbf{m},2}^{(2)\mathrm{c}} - \Tilde{M}_{\textbf{m},0}^{(2)} - \frac{1}{2} \Tilde{M}_{\textbf{m},4}^{(2)\mathrm{c}} \right) \right] \cos{(2 \omega t)}\, \partial^{\textbf{m}} \\
        &+ \frac{1}{2} \left[ 2 \Tilde{M}_{\textbf{m},2}^{(0)\mathrm{s}} + \alpha_0 \left( - \Tilde{M}_{\textbf{m},3}^{(1)\mathrm{c}} + \Tilde{M}_{\textbf{m},1}^{(1)\mathrm{c}} \right) + \alpha_0^2 \left( - \Tilde{M}_{\textbf{m},2}^{(2)\mathrm{s}} + \frac{1}{2} \Tilde{M}_{\textbf{m},4}^{(2)\mathrm{s}} \right) \right] \sin{(2\omega t)}\, \partial^{\textbf{m}}\ , \\[3mm]
    \left( \hat{\Omega} \Delta \hat{\Omega}^{\dagger} \right)_n =\, &\frac{1}{2} \left[ 2 \Tilde{M}_{\textbf{m},n}^{(0)\mathrm{c}} + \alpha_0 \left( - \Tilde{M}_{\textbf{m},n+1}^{(1)\mathrm{s}} + \Tilde{M}_{\textbf{m},n-1}^{(1)\mathrm{s}} \right) + \alpha_0^2 \left( \Tilde{M}_{\textbf{m},n}^{(2)\mathrm{c}} - \frac{1}{2} \left( \Tilde{M}_{\textbf{m},n+2}^{(2)\mathrm{c}} + \Tilde{M}_{\textbf{m},n-2}^{(2)\mathrm{c}} \right) \right) \right] \cos{(n \omega t)}\, \partial^{\textbf{m}} \\
        &+ \frac{1}{2} \left[ 2 \Tilde{M}_{\textbf{m},n}^{(0)\mathrm{s}} + \alpha_0 \left(- \Tilde{M}_{\textbf{m},n+1}^{(1)\mathrm{c}} + \Tilde{M}_{\textbf{m},n-1}^{(1)\mathrm{c}} \right) + \alpha_0^2 \left( - \Tilde{M}_{\textbf{m},n}^{(2)\mathrm{s}} + \frac{1}{2} \left( \Tilde{M}_{\textbf{m},n+2}^{(2)\mathrm{s}} + \Tilde{M}_{\textbf{m},n-2}^{(2)\mathrm{s}} \right) \right) \right] \sin{(n \omega t)}\, \partial^{\textbf{m}}\ .
    \label{eq:harmonic_laplace}
\end{aligned}
\end{align}
\end{widetext}
The time-independent part yields the operator $\Delta_{0} \coloneqq \left( \hat{\Omega} \Delta \hat{\Omega}^{\dagger} \right)_0 $ that is inserted in the transformed Schrödinger equation within the main text, while the other terms are collected into the perturbative term $\delta\Delta (t)$.

\FloatBarrier
\bibliography{KH_I}

\end{document}